\documentstyle[mprocl]{article}
\begin{document}
\begin{flushright}
gr-gc/9709056
\end{flushright}
\vskip 0.3cm
\title{ON THE STABILITY OF REAL SCALAR BOSON STARS}

\author{PHILIPPE JETZER}
\address{Paul Scherrer Institute, Laboratory for Astrophysics, CH-5232 
Villigen PSI, and
Institute of Theoretical Physics, University of Z\"urich, Winterthurerstr.
190, CH-8057 Z\"urich, Switzerland}

\author{ DAVID SCIALOM} 
\address{D\'epartement d'Astrophysique Relativiste et de Cosmologie,
\\CNRS-Observatoire de Paris, 92195 Meudon, France}

\maketitle

\abstracts{We discuss spherically symmetric static
solutions of the Einstein-Klein-Gordon 
equations for a real scalar field with a mass and a quartic
self-interaction term. As for the massless case the solutions have a naked
singularity at the origin. However, linear stability analysis shows that 
these solutions
as well as the massless ones are dynamically unstable.}

\section{Introduction}

The recent developments in particle physics and cosmology suggest that 
scalar fields may have played an important role in the evolution of the
early universe, for instance in primordial phase transitions, and
that they may make up part of the dark matter.
These facts motivated the study of gravitational equilibrium
configurations of scalar fields, in particular for massive
complex fields, which form so-called boson stars
(for a review on this subject see Refs. 1 and 2).
%\cite{Jet:92,Lid:92}).
Solutions to the Einstein-Klein-Gordon equations for massless real
scalar fields were first studied by Buchdal \cite{bu:59} and
more recently by Wyman \cite{wy}.
All known static solutions for the real scalar field to the 
Einstein-Klein-Gordon equations are either 
topologically non-trivial or have a logarithmic singularity. 
The latter is an 
example of a so-called naked singularity.   

Here we present the static, spherically symmetric
solutions for a real scalar field obtained by including 
a mass and a quartic self-interaction term.
We also discuss the dynamical stability of the equilibrium solutions
using linear perturbation theory. 
Using the variational principle, from which the
pulsation equation is derived, we find for all equilibrium solutions
a negative upper bound
for the lowest eigenvalue. We, therefore, conclude that all equilibrium
configurations, 
including the massless ones, are dynamically unstable.

\section{Basic equations and their properties}

We consider the action of a real scalar field $\phi$ with mass and
a quartic self-interaction term ($\tilde{\lambda}>0$) minimally
coupled to gravity with the following spherically symmetric metric 
\begin{equation}
ds^2=e^{\nu}dt^2-e^{\lambda}dr^2-r^2(d\vartheta^2+sin^2\vartheta d\varphi^2)~.
\label{eq:b}
\end{equation}
The static equilibrium configurations are determined by a
coupled system of equations, which can be derived from the action,
by varying with respect to the various fields.
For all solutions we require as boundary conditions that the metric is 
asymptotically flat and that $\phi$ vanishes at infinity. 
It turns out that at the origin
$\phi$ is singular and behaves for all cases as: $\phi \sim -ln~r$.
For $r \rightarrow 0$: $e^{\lambda} \sim r^a$ ($a$ is real and $a>1$) 
and $e^{\lambda} \rightarrow 0$, whereas  
$e^{\nu}\sim r^b$ ($b$ is real and $b>-1$). 
For $e^{\nu}$ there are three different possible 
behaviors: $e^{\nu} \equiv 1$, $e^{\nu}\rightarrow \infty$ or 
$e^{\nu} \rightarrow 0$. See Ref. 5
%\cite{Sci:92} 
for a plot of the results of
the numerical integration for $\phi$ and the metric functions.
The corresponding ADM mass of the boson
star has been also computed numerically. 

\section{Dynamical stability}

We turn now to the problem of dynamical stability and consider small
time dependent
radial perturbations, which still preserve spherical symmetry. The equations
governing the linear perturbations are obtained by expanding all functions
to first order ($\lambda(r,t)=\lambda_0(r)+\delta\lambda(r,t);
\nu(r,t)=\nu_0+\delta\nu(r,t); \phi(r,t)=\phi_0(r)+\delta\phi(r,t)$)
and by linearizing the Einstein-Klein-Gordon equations. We make
also use of the $G_{01}$ component of the 
Einstein equation, which can be integrated once in
time \cite{Jet:89}. 
This way the metric functions $\delta\lambda$ and $\delta\nu$ can
be eliminated from the linearized scalar wave equation 
\cite{Jet:89,Jet:92}.
Furthermore, we suppose a time dependence of the form $e^{i\sigma t}$ for
$\delta\phi(r,t)$. Thus $\delta\phi(r,t)=e^{i\sigma t} \Psi(r)/r$, where
we denote the radial part of $\delta\phi$ by $\Psi(r)/r$.
Performing a change of variable defined by
$\frac{d\rho}{dr}=e^{(\lambda_0-\nu_0)/2}$ with $\rho(r=0)=0$, the
pulsation equation transforms to the following Schr\"odinger-type equation
\begin{equation}
-\frac{d^2 \Psi}{d\rho^2}+V(r[\rho])\Psi=\sigma^2 \Psi~, \label{eq:o}
\end{equation}
where 
\begin{eqnarray}
V(r)&=&e^{\nu_0-\lambda_0}((\nu_0^{\prime}-\lambda_0^{\prime})/
2r 
-4\pi G r \phi_0^{\prime 2}(2/r+\nu_0^{\prime}-\lambda_0^{\prime})
\nonumber \\ & &
+16\pi G r \phi_0^{\prime} e^{\lambda_0}(m^2\phi_0+\tilde\lambda\phi_0^3)+
e^{\lambda_0}(m^2+3\tilde\lambda\phi_0^2))   \label{eq:p}       
\end{eqnarray}
($r=r(\rho)$). 
Dynamical instability occurs whenever the lowest eigenvalue $\sigma_0^2$
is negative. Indeed, the perturbation $\delta\phi \sim e^{i\sigma_0 t}$ 
will then grow exponentially. Since the asymptotic behavior of the solution
$\phi_0(r)$ does not depend on the mass nor on the quartic
self-interaction term, the behavior of
$V(\rho)$ will not depend
on $m$ and $\tilde\lambda$ and thus will be the same for all types
of solutions. One finds $V(\rho)\sim -\frac{1}{4\rho^2}$ for 
$\rho \rightarrow 0$. 
It is useful to write the potential $V(\rho)$ as:
$V(\rho)=-\frac{1}{4\rho^2}+\tilde V(\rho)$. 
This way we can write the Hamilton operator as $H=H_0+\tilde V(\rho)$,
where $H_0=-\frac{d^2}{d\rho^2}-\frac{1}{4\rho^2}$.
$H_0$ is not selfadjoint on its natural domain of definition 
${\cal D}(H_0)={\cal D}(\frac{d^2}{d\rho^2})\cap{\cal D}(\frac{1}
{4\rho^2})$, since ${\cal D}(H_0)\neq {\cal D}(H_0^{*})$.
This problem can be solved by extending in an appropriate way the 
domain of definition of $H_0$, such that for the corresponding operator
$H_{0,a}$~ ($a$ is real)~  
${\cal D}(H_{0,a})\subset {\cal D}(H_{0}^{*})$ and ${\cal D}(H_{0,a})=
{\cal D}(H_{0,a}^{*})$.
We define \cite{Nar:74}
${\cal D}(H_{0,a})={\cal D}(H_0)+\Psi_a$ with
\begin{equation} 
\Psi_a=\sqrt{\rho}(H_0^{(1)}[exp(i\frac{\pi}{4})\rho]
+exp(ia)\overline{H_0^{(1)}  
[exp(i\frac{\pi}{4})\rho]})~,    \label{eq:cc}
\end{equation} 
and 
$H_0^{(1)}$ is the Hankel function. The spectrum of $H_{0,a}$
depends on the value of $a$,
which parameterizes the extension. For our problem 
$\Psi_a$ must be real, since it describes a perturbation of the equilibrium 
solution for a real scalar field.
$\Psi_a$ is real only for $a=0$.
On this domain of definition ${\cal D}(H_{0,a=0})$ $H_0$ 
(means $H_{0,a=0}$ from now on) has a discrete 
eigenvalue $E=0$, whose eigenfunction $\Psi_{0}$=$\Psi_{a=0}$ is in 
${\cal L}^2(d\rho,(0,\infty))$.
From the scaling property of $H_0$, it follows that this eigenvalue is
infinitely degenerated.

Using $\Psi_{0,\beta}(\rho)$=$\Psi_0(\rho\beta)$   
as a trial function we get upper bounds for the eigenvalue $\sigma_0^2$:
\begin{equation}
\sigma_0^2 \leq \frac{<\Psi_{0,\beta} \mid H_0+\tilde V \mid \Psi_{0,\beta}>}
{<\Psi_{0,\beta} \mid \Psi_{0,\beta}>}=\frac{<\Psi_{0,\beta} \mid \tilde V 
\mid \Psi_{0,\beta}>}{<\Psi_{0,\beta} \mid \Psi_{0,\beta}>}~.   \label{eq:dd}
\end{equation}
For all equilibrium solutions
we easily find a value for $\beta$ such that 
$<\Psi_{0,\beta}\mid \tilde V\mid \Psi_{0,\beta}>$ is negative.
Indeed, for large $\beta$, $\Psi_0 (\beta\rho)$ reaches its maximum at
smaller values of $\rho$ than for $\beta=1$. 
$\tilde V(\rho)$ is negative at the origin and
behaves as $-\frac{const}{\rho}$ for $\rho \rightarrow 0$.
Thus by appropriately choosing $\beta$, $\Psi_0(\beta\rho)$ has most of its
support where $\tilde V(\rho)$ 
is negative, which leads then to a negative
value for $<\Psi_{0,\beta}\mid\tilde V\mid\Psi_{0,\beta}>$.
Therefore, we conclude
that all the static solutions are dynamically unstable \cite{Sci:92}.
This result is in agreement with the cosmic censorship conjecture 
\cite{Wal:84}, which
excludes spacetimes with naked singularities.   
\section{Conclusion}
From the above considerations it follows that
only with complex scalar
fields one may hope to 
form boson stars, and it is thus natural to study their formation
and properties within
the standard cosmological model. 
To that purpose an analysis of the coupled Einstein-Klein-Gordon
equations using the Friedmann-Lema\^{\i}tre metric has
been carried out for a complex field 
in Ref. 9.
%\cite{sci:95}. 
Moreover, the time evolution of the perturbations of 
the complex scalar field and the metric has been analyzed
in Ref. 10.
% \cite{sci:97}. 
It turns out, that during the
oscillatory phase after inflation the
perturbations of the complex scalar field at best oscillate.
Therefore, the formation of boson stars in a universe driven by the same
scalar field is not possible. \\

D.S. is supported by the Swiss National Science Foundation.

%\bibliography{total}
%\bibliographystyle{WorldScientific}    % for BibTeX - sorted numerical
                                       % labels by order of    
                                       % first citation. 
\end{document}